\documentclass[12pt,a4paper,final]{iopart}
\usepackage{iopams}
\usepackage{float}
\usepackage{graphicx}
\usepackage[breaklinks=true,colorlinks=true,linkcolor=blue,urlcolor=blue,citecolor=blue]{hyperref}
\usepackage{tabularx}

\begin{document}

\title{Magnetotransport of single crystalline NbAs}

\author{N.J. Ghimire, Yongkang Luo, M. Neupane, D.J. Williams, E.D. Bauer, and F. Ronning}

\address{Los Alamos National Lab, Los Alamos, NM, 87544, USA}

\ead{fronning@lanl.gov}

\begin{abstract}
We report transport measurement in zero and applied magnetic field on a single crystal of NbAs. Transverse and longitudinal magnetoresistance in the plane of this tetragonal structure does not saturate up to 9 T. In the transverse configuration ($H \parallel c$, $I \perp c$) it is 230,000 \% at 2 K. The Hall coefficient changes sign from hole-like at room temperature to electron-like below $\sim$ 150 K. The electron carrier density and mobility calculated at 2 K based on a single band approximation are 1.8 x 10$^{19}$ cm$^{-3}$ and 3.5 x 10$^{5}$ cm$^2$/Vs, respectively. These values are similar to reported values for TaAs and NbP, and further emphasize that this class of noncentrosymmetric, transition-metal monopnictides is a promising family to explore the properties of Weyl semimetals and the consequences of their novel electronic structure.
\end{abstract}

\vspace{2pc}

\section{Introduction}
Over the past few years, it has been appreciated that the Berry curvature (or Chern flux) of filled bands in an insulator can lead to new topological classifications of matter \cite{Qi2011,Hasan2010}. More recently, it was noticed that the Berry curvature also results in a new topological classification of metallic systems. The Weyl semimetal has sinks and sources of Berry curvature in momentum space where the band structure forms Dirac cones \cite{Balents2011,Hosur2013}. Due to the electronic dispersion, Weyl semimetals can be considered as 3D analogs to graphene. One of the most striking consequences of a Weyl semimetal is that the surface possesses Fermi arcs \cite{Hosur2012,Ojanen2013,Potter2014}. The bulk may also possess novel signatures in transport properties as a consequence of the chiral anomaly \cite{Zyuzin2012,Wang2013, Parameswaran2014}. A necessary condition to create a Weyl semimetal is to break either time reversal symmetry or inversion symmetry. Though magnetic ordering has been employed to predict the existence of Weyl semimetals \cite{Xu2011,Burkov2011,Wan2012}, magnetism is, in general, a difficult property to predict from first principles calculations. A more straightforward approach is to examine uncorrelated systems, which lack inversion symmetry, for which density functional theory calculations employing the local density (or generalized gradient) approximation provide accurate descriptions of the electronic structure.

Using this approach the transition-metal monopnictides, TaAs, TaP, NbAs, and NbP, have recently been predicted to be Weyl semimetals \cite{Hongmin2015, Huang2015}. All four compounds crystallize in the noncentrosymmetric NbAs structure-type with the I4$_{1}md$ space group \cite{Furuseth1964}. Angle-resolved photoemission results report the observation of the predicted Fermi arc states in TaAs \cite{Xu2015,Lv2015}. Transport studies on TaAs reveal a gigantic magnetoresistance of 540,000 \% at 10 K and 9 T, which is non-saturating up to 56 T, large mobilities of 5 x 10$^5$ cm$^2$/Vs, and a negative magnetoresistance in the longitudinal configuration \cite{Zhang2015,HuangX2015}, the latter of which has been interpreted, as in Bi$_{1-x}$Sb$_{x}$ \cite{Kim2013}, as evidence for the anticipated chiral anomaly. Subsequently, a single crystal study of NbP showed similar properties (850,000 \% magnetoresistance at 2 K and 9 T with a mobility of 5 x 10$^6$ cm$^2$/Vs) \cite{Shekhar2015}. Here we report transport measurements on single crystals of NbAs. We find very similar results to TaAs and NbP with regards to large mobilities and the magnetoresistance which reaches 230,000 \% at 2 K and 9 T, and is growing. However, similar to NbP, no negative longitudinal magnetoresistance was observed over this field and temperature range.

\section{Methods}
 Single crystals of NbAs were grown by vapor transport using iodine as the transport agent. First, polycrystalline NbAs was prepared by heating stoichiometric amounts of Nb and As in an evacuated silica ampoule at 700 $^{\circ}$C for 3 days. Subsequently, the powder was loaded in a horizontal tube furnace in which the temperature of the hot zone was kept at 950 $^{\circ}$C and that of the cold zone was $\approx$ 850 $^{\circ}$C. Several NbAs crystals formed with distinct well facetted flat plate like morphology. The crystals of NbAs were verified by checking (00l) reflections on a x-ray diffractometer and by compositional analysis conducted using an energy dispersive x-ray spectroscopy (EDS). An atomic percentage ratio of Nb:As = 53:47 was obtained on the EDS measurements, which is close to the expected uncertainty of 3-5 \%.   A plate-like crystal with thickness $t$ = 68 $\mu$m was selected for the transport studies. Electrical contacts were made by spot welding 25 $\mu$m Pt wires onto the sample in a Hall bar geometry, with length of 0.6 mm and width of 1.2 mm. An ac-resistance bridge was used to perform the transport measurements in a Quantum Design PPMS. Longitudinal (transverse) voltages were symmeterized (antisymmeterized) about $H$ = 0 T, to account for the small misalignment of the voltage leads.

\section{Crystal Chemistry}
\begin{figure}
\begin{center}
\includegraphics[width=20pc]{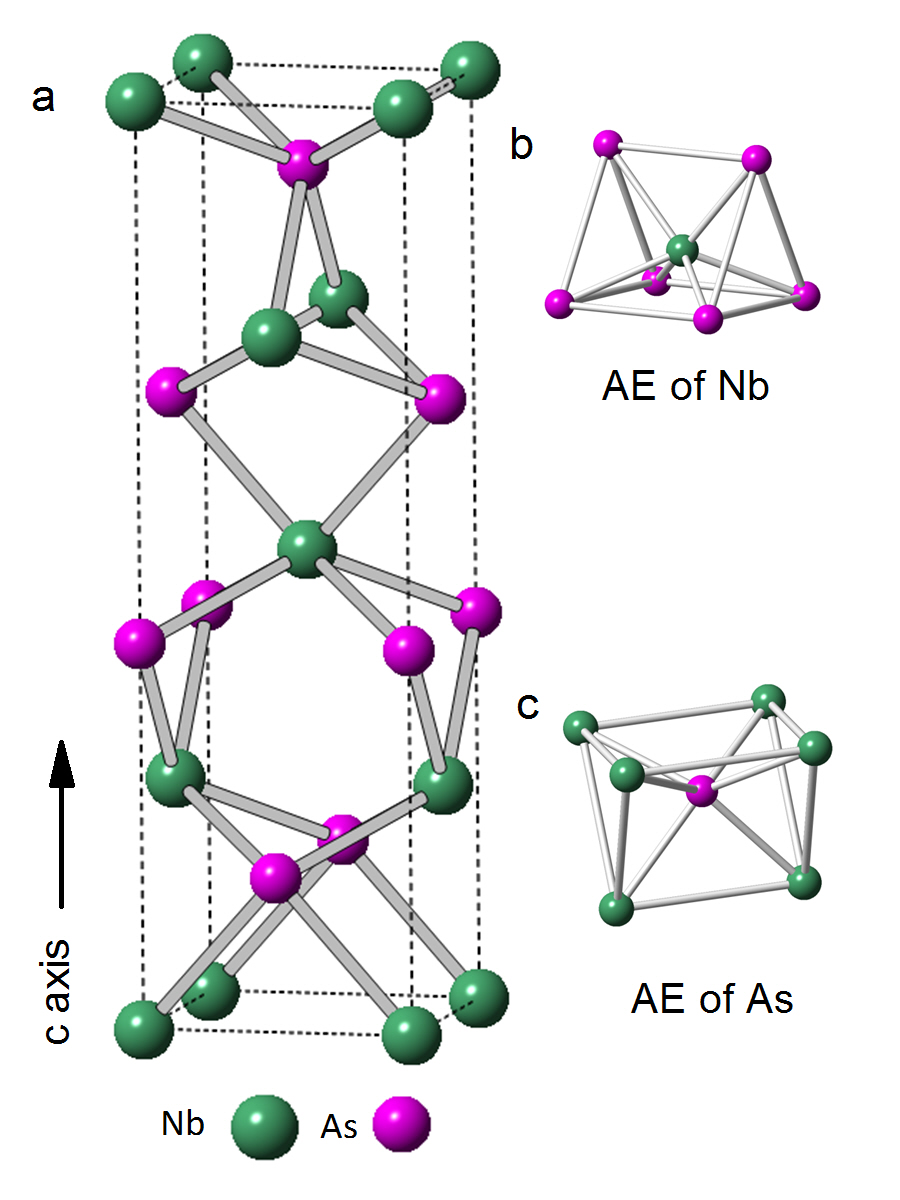}
\caption{\label{Fig1a} (a) Crystal structure of NbAs. (b,c) Trigonal prism atomic environment (AE) of Nb, As.}
\end{center}
\end{figure}

NbAs crystallizes in the noncentrosymmetric tetragonal space group I4$_{1}$md (\# 109) and is isostuctural to TaAs, TaP and NbP \cite{Furuseth1964}. The structure is made up of a three dimensional network of trigonal prisms of NbAs$_{6}$ and AsNb$_{6}$ as shown in Fig. \ref{Fig1a}. There is a single site for both Nb and As atoms in the unit cell. Both Nb and As occupy the 4a Wyckoff position. The atomic coordinates of Nb and As are (0, 0, 0) and (0, 0, 0.416), respectively and the reported lattice parameters are a = 3.4517 {\AA} and c = 11.1680 {\AA} \cite{Furuseth1964}.

\section{Results and Discussion}
\begin{figure}
\begin{center}
\includegraphics[width=40pc]{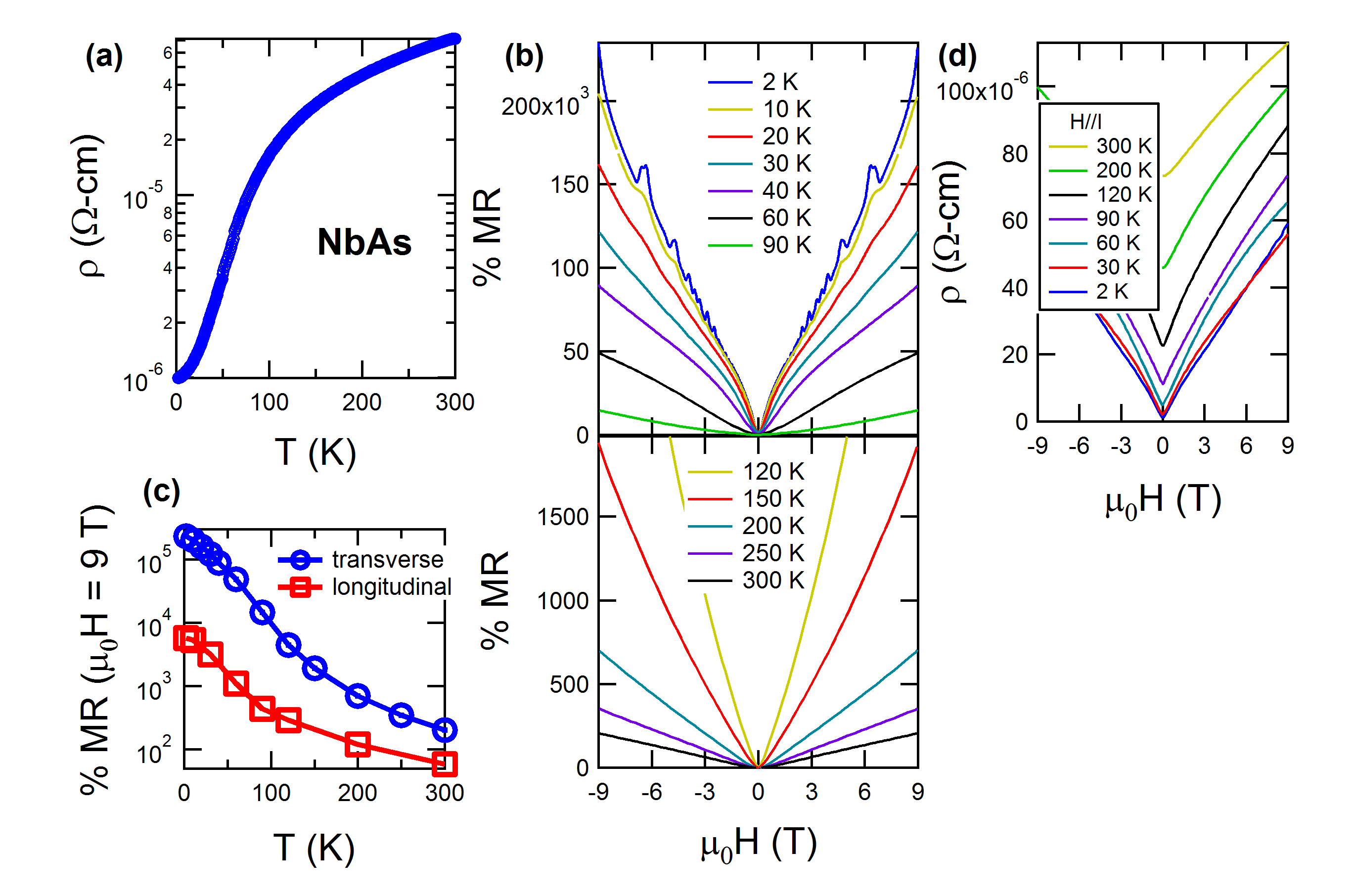}
\caption{\label{Fig1} (a) Temperature dependent, zero field, in-plane resistivity of NbAs. (b) Transverse in-plane \% MR (= 100 x $(\rho_{xx}(\mu_{0}H) - \rho_{xx}(\mu_{0}H=0))/\rho_{xx}(\mu_{0}H=0)$) versus $H$ applied along the $c$-axis at different temperatures. (c) Transverse (circles) and longitudinal (squares) \% MR at 9 T with $H\parallel c$. (d) Longitudinal ($I  \parallel H; I \perp c$) in-plane $\rho$ versus $H$ at different temperatures.}
\end{center}
\end{figure}

The zero field in-plane resistivity of NbAs is shown in Fig. \ref{Fig1}a. The resistivity decreases with decreasing temperature as expected for a metallic system and in agreement with previous reports on polycrystalline samples\cite{Saparov2012}. The residual resistivity ratio (RRR = $\rho$(300K)/$\rho$(2K)) is 72, indicative of the high sample quality of these single crystals. Transverse magnetoresistance defined as \% MR = 100 x $(\rho_{xx}(\mu_{0}H) - \rho_{xx}(\mu_{0}H=0))/\rho_{xx}(\mu_{0}H=0)$ with $H\parallel c$ and $I \perp c$ is presented in Fig. \ref{Fig1}b. At room temperature the magnetoresistance is approximately linear and reaches 200 \% at 9 T. Upon cooling the magnetoresistance increases dramatically with a nearly exponential rise with decreasing temperature (see Fig. \ref{Fig1}c). At the lowest temperature of 2 K the magnetoresistance is over 230,000 \%.  This value is similar (to within a factor of 3) to what was observed in TaAs and NbP in samples with comparable RRR \cite{Zhang2015,HuangX2015,Shekhar2015}. Large magnetoresistance is a signature of compensated metals with high mobilities. Attempts to scale the MR data with a Kohler plot MR = $F(H/\rho_{xx}(H=0,T))$ fails (not shown) indicating that there are multiple scattering rates, which is not unexpected for a multiband system. In addition to the large magnetoresistance, additional features become stronger with decreasing temperature that are consistent with Shubnikov-de Haas oscillations. However, no single frequency could be identified, and we leave an analysis of the Fermi surface topology for future work.

One consequence of the chiral anomaly which exists in a Weyl semimetal is that the magnetoresistance should have an additional negative contribution when current is parallel to the magnetic field \cite{Hosur2013}. Rotating the magnetic field into the plane can also suppress the large positive contribution to the magnetoresistance. In Fig. \ref{Fig1}d the longitudinal magnetoresistance of NbAs is plotted. At the base temperature of 2 K it remains large and positive, though as illustrated in Fig. \ref{Fig1}c it is more than an order of magnitude smaller than the MR observed in the transverse configuration. An examination of the data at low magnetic fields does not reveal any clear indication of a negative magnetoresistance contribution.

\begin{figure}
\begin{center}
\includegraphics[width=30pc]{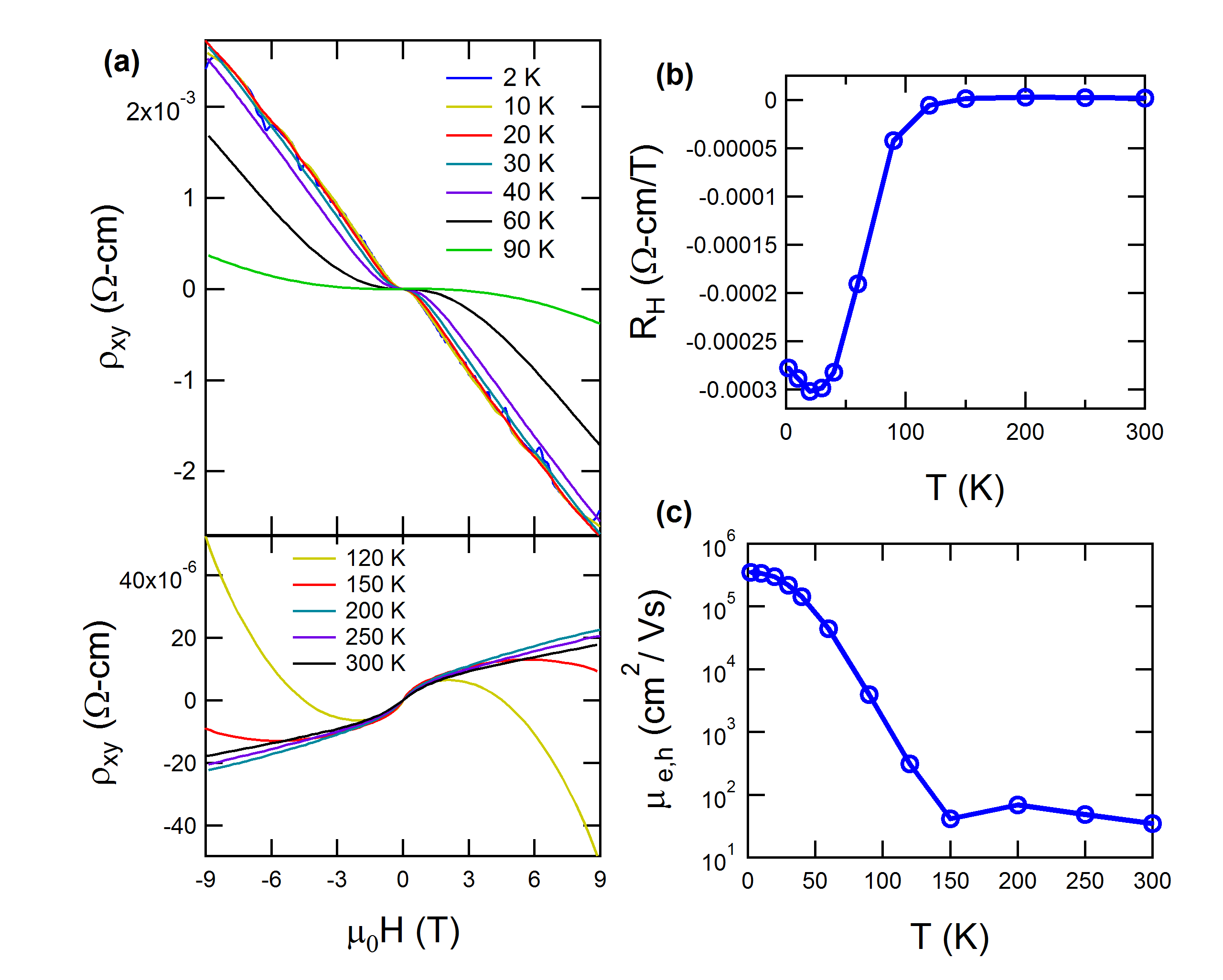}
\caption{\label{Fig2} (a) In-plane Hall resistivity of NbAs with magnetic field applied along the $c$-axis. (b) Hall coefficient $R_H$ at 9 T as a function of temperature. (c) Mobility versus temperature determined by the Hall coefficient at 9 T and the zero field resistivity using a single band approximation.}
\end{center}
\end{figure}

The Hall resistivity is $\rho_{xy}$ = $V_y$/$I_x$$t$, where $V_y$ is the transverse voltage which develops in the presence of a magnetic field along the $c$-axis with current $I_x$ in the $ab$-plane. Figure \ref{Fig2}a plots the Hall resistivity as a function of magnetic field and for different temperatures. At high temperatures the field dependence is positive indicative of hole-like carriers, while at low temperatures the Hall resistivity changes sign and grows significantly. This is consistent with multiple hole- and electron-like carriers as was also observed in TaAs and NbP, and indicated by band structure calculations \cite{Hongmin2015,Huang2015,Zhang2015,HuangX2015,Shekhar2015}. To estimate the carrier concentration and the mobility we analyze the data within a single band model. This is most applicable at low temperatures where the electron contribution dominates the Hall resistivity.  The carrier concentration is given by $n$ = 1/$eR_H$, where $e$ is the charge of the electron and we take the Hall coefficient $R_H$ equal to $\rho_{xy}$/$\mu_{0}H$ at 9 T. The mobility is $\mu$ = 1/$en\rho_{xx}$($\mu_{0}H$=0). The temperature dependent data for $R_H$ at 9 T and $\mu$ is shown in Figs. \ref{Fig2}b and \ref{Fig2}c, respectively. At 2 K we find a carrier concentration of 1.8 x 10$^{19}$ cm$^{-3}$, which is an upper bound, as any hole contribution would artificially inflate the computed carrier concentration under the single band approximation. From the carrier concentration we obtain a mobility at 2 K of the electron carriers of 3.5 x 10$^{5}$ cm$^2$/Vs. Again, this is similar to the value of 5 x 10$^{6}$ cm$^2$/Vs obtained in NbP under a single band approximation\cite{Shekhar2015}. As temperature increases the computed carrier concentration grows as the hole carriers increasingly compensate the electron contribution, and the mobility decreases.

\section{Conclusion}
In summary, we have grown single crystals of the proposed Weyl semimetal NbAs using vapor transport. The electrical transport properties are overall very similar to previously reported studies of TaAs and NbP \cite{Zhang2015,HuangX2015, Shekhar2015}. The magnetotransport displays a huge magnetoresistance, which does not saturate up to 9 T, with strikingly large mobilities at 2 K. These results make NbAs an intriguing candidate worthy of further study. At the moment it is unclear whether the dramatic transport properties of NbAs are a consequence of the non-trivial topology of its electronic structure, which define it as a Weyl semimetal, or whether it is simply a clean semimetal such as WTe$_2$ or Bi \cite{Ali2014,Yang1999}, with no influence of peculiar topological band structure effects. In this regard, it may be interesting that the longitudinal magnetoresistance in NbAs (our work) and NbP \cite{Shekhar2015} remains positive, while TaAs has reported negative longitudinal magnetoresistance attributed to the chiral anomaly \cite{Zhang2015, HuangX2015}. Spin-orbit coupling, which is required to produce the Weyl points will be much weaker in Nb relative to Ta. Hence, the negative magnetoresistance may be directly linked to the influence of the spin-orbit coupling on the electronic structure. Alternatively, the Fermi energy may be located further from the Weyl points in the Nb compounds relative to TaAs.

\subsection{Acknowledgments}
We thank Joe Thompson for useful discussions. Samples were synthesized under the auspices of the Department of Energy, Office of Basic Energy Sciences, Division of Materials Science and Engineering. The EDS measurements were performed at the Center for Integrated
Nanotechnologies, an Office of Science User Facility operated
for the US Department of Energy (DOE) Office of Science.
Los Alamos National Laboratory, an affirmative action
equal opportunity employer, is operated by Los Alamos
National Security, LLC, for the National Nuclear Security
Administration of the US Department of Energy under contract
DE-AC52-06NA25396. Electrical transport measurements were supported by the LANL LDRD program.


\section*{References}
\bibliographystyle{iopart-num}

\end{document}